# Prediction of two-dimensional ferromagnetic ferroelectric VOF$_2$ monolayer

Hai-Peng You, Ning Ding, Jun Chen, and Shuai Dong*

School of Physics, Southeast University, Nanjing 211189, China

**Abstract**

Nowadays, designing and searching for materials with multiple functional characteristics are the key to achieving high-performance electronic devices. Among many candidates, two-dimensional multiferroic materials have great potential to be applied in highly-integrated magnetoelectric devices, such as high-density non-volatile memories. Here, we predict a two-dimensional material VOF$_2$ monolayer to possess intrinsic ferroelectric and ferromagnetic properties. The VOF$_2$ monolayer own the largest in-plane ferroelectric polarization (332 $p$C/m) in the family of VO$X_2$ ($X$: halogen) oxyhalides. Different from other VO$X_2$ monolayers whose magnetic ground states are antiferromagnetic or noncollinear spiral textures, the VOF$_2$ monolayer owns a robust ferromagnetic ground state, which is rare but highly desirable. Our theoretical prediction provides a good candidate and starting point for the further pursuit of more two-dimensional multiferroic materials with high-performance magnetoelectricity.

*Corresponding Author. E-mail: sdong@seu.edu.cn

# Introduction

In the last decade, the researches on two-dimensional (2D) van der Waals (vdW) materials have developed rapidly for their broad application prospects in the field of microelectronics and optelecronics. But untill very recent years, the 2D materials with intrinsic ferro-type properties, like ferroelectricity and ferromagnetism, emerged as attractive new branches of 2D families, such as $CuInP_2S_6$,[1-4] $In_2Se_3$,[5,6] $SnTe$,[7] $Cr_2Ge_2Te_6$,[8] and $CrI_3$.[9,10] In three-dimensional (3D) crystals,[11] the ferro-materials are high-valuable functional materials,[12-14] which have broadly served in many devices. The 2D forms of these ferro-ordered materials can be even more interesting and useful, considering the urgent demanding of miniaturization for devices.

Encouraged by the recent success in 2D ferroelectric (FE) materials and ferromagnetic (FM) materials,[15-19] the investigations on 2D multiferroics are also starting.[20,21] Multiferroics, coupling ferroelectricity and magnetism in single phase materials, can be much helpful to realize ultra-high speed reading and writing for data storages.[22-26] Thanks to the great progresses achieved in the past decades, the physical understanding of multiferroic materials have been much deepen. Even through, in 3D crystals, almost all known multiferroics are antiferromagnetic (AFM), while the desired materials with intrinsic ferromagnetism and ferroelectricity remain rare. Fortunately, recently Tan *et al.* predicted a series of 2D multiferroic oxyhalides $VOX_2$ ($X$=Cl, Br, and I), among which the magnetic ground state of $VOI_2$ was predicted to be FM.[27] In contrast, $VOCl_2$ and $VOBr_2$ were proposed to be AFM, as confirmed by Ai *et al.*[28]

However, our following study found that $VOI_2$ could not possess ferromagnetism and ferroelectricity simultaneously. The crucial reason is that the strong spin-orbit coupling (SOC) of heavy element iodine can lead to a large Dzyaloshinskii-Moriya interaction for the polar structure, which distorts the FM texture to a short-periodic spiral one. Thus, heavy elements should be avoided if one wish to stabilize ferromagnetism in ferroelectric systems. Along this way, we will study the end member in the opposite side, i.e. $VOF_2$, which was missed in previous studies,[27,28] to pursue the possible coexistence of ferromagnetism and ferroelectricity.

# Computational methods

The density of functional theory (DFT) calculations are performed using the Vienna *ab initio* Simulation Package (VASP).[29,30] The projector-augmented wave (PAW) potentials with generalized gradient approximation of Perdew-Burke-Ernzerhof (GGA-PBE)[31] formulation are used with a cut-off energy of 550 eV. In order to search for possible magnetic ground state of $VOF_2$, a 2×2×1 supercell is constructed. A 9×9×1 $k$-grid is adopted to sample the Brillouin zone utilizing the Monkhorst-Pack $k$-points scheme for the unit cell.[32] The energy convergence criterion is $10^{-6}$ eV and a criterion of Hellman-Feynman force is 0.005 eV/Å throughout the structural relaxation. The FE polarization is obtained by the standard Berry phase methods.[33] A vacuum layer of 16 Å is added in the out-of-plane direction to avoid interactions between adjacent layers. The vdW interaction was described by the DFT-D2 functional.[34] The DFPT+Phonopy method is employed to calculate phonon spectra.[35] The GGA+$U$ approach is adopted for $3d$ orbitals of transition metal V.[36] For comparison, the hybrid functional calculations based on the Heyd-Scuseria-Ernzerhof (HSE06) exchange are also done.[37] According to literature, the coefficient AEXX=0.1 for V ion is a proper choice.[38] To analyze the thermodynamic properties, the Markov-chain Monte Carlo (MC) method with Metropolis algorithm was employed to simulate the magnetic phase transition under finite temperature. The MC simulation was done on a 36×36 lattice with periodic boundary conditions and larger lattices were also tested to confirm the physical results. The initial $2\times10^4$ MC steps were discarded for thermal equilibrium and the following $2\times10^4$ MC steps were retained for statistical averaging of the simulation. To avoid the possible trapping in local minimal energy, the quenching process was used for the temperature scanning. To characterize the magnetic phase transitions, the specific heat was calculated to determine the critical temperature.

**Results and discussions**

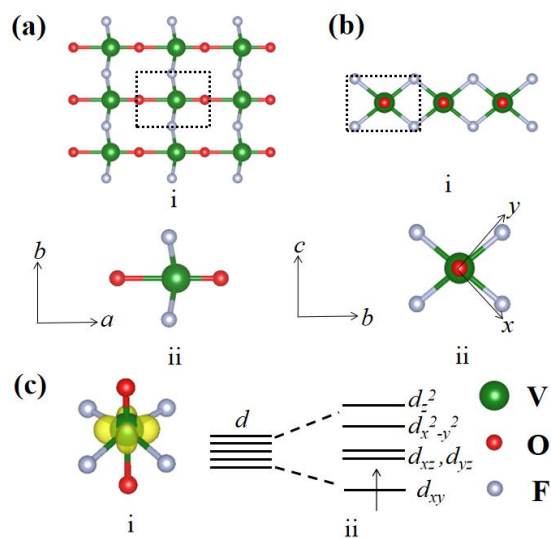

**Figure 1.** Schematic of $VOF_2$ structure, which is similar to other $VOX_2$. (a) Top view of $VOF_2$

monolayer. i: a 3×3 supercell; ii: a minial unit cell. (b) Side view along the *a*-direction. i: a 3×1 supercell; ii: a minial unit cell. The dashed rectangles in (a)i and (b)i indicate the primitive cells. The *x-y* axis of octaheron is also indicated in (b)ii. (c) The orbital structure. Left: the charge density of occupied *d*-orbital. Right: the orbital splitting due to the crystal field.

**Table I.** DFT optmized structural parameters of VOF$_2$ monolayer by PBE and HSE06 methods, which are quite close to each other.

|  | Lattice constants | Bond lengths | Bond angles |
| --- | --- | --- | --- |
| **PBE** | $a$=3.783Å | 1.973Å(V-F) | 180°(O-V-O) |
|  | $b$=3.033Å | 2.152Å (V-O$_1$) | 82.5°(F-V-O$_1$) |
|  |  | 1.630Å (V-O$_2$) | 97.5°(F-V-O$_2$) |
| **HSE06** | $a$=3.738 Å | 1.954 Å (V-F) | 180° (O-V-O) |
|  | $b$=3.013 Å | 2.133 Å (V-O$_1$) | 82.7°(F-V-O$_1$) |
|  |  | 1.604 Å (V-O$_2$) | 97.3° (F-V-O$_2$) |

**Structure and stability**

Since the experimental data are available only for oxyhalide VOCl$_2$ bulk,[39] the initial structure of VOF$_2$ monolayer, can be obtained by replacing Cl with F. Considering the fact that element F is more electronegative than Cl, VOF$_2$ should be more chemically stable than VOCl$_2$ against oxygen and moisture.

Then the structural relaxation is performed. As shown in Fig. 1(a), the crystal structure of VOF$_2$ monolayer is orthorhombic (space group No. 25 *Pmm*2) after optimization. The V ion, caged in the O$_2$F$_4$ octahedron, shifts towards one oxygen ion, which breaks the spatial inversion symmetry along the *a*-axis and leads to a polar structure. The optimized lattice structural parameters of VOF$_2$ monolayer by PBE and HSE06 can be found in Table I. Comparing with other oxyhalides VO*X*$_2$ (*X*=Cl, Br, and I),[27] VOF$_2$ has a smaller *b* due to smaller radius of F$^{1-}$ than other halogen ions.

Due to the anisotropic octahedral crystal field, the 3*d* orbitals of V ion are splitted into four energy levels: $d_{z^2}$, $d_{x^2-y^2}$, $d_{xz}/d_{yz}$, and $d_{xy}$ orbitals, respectively. The $d_{xy}$ orbital has the lowest energy, and thus the single unpaired electron of V$^{4+}$ should occupy this $d_{xy}$ orbital, as shown in

Fig. 1(c). For most proper ferroelectrics, the empty $d$ orbitals are preferred, i.e. the emprical $d^0$ rule. Here, although the electronic structure is not $d^0$ but $d^1$, the orbitals $d_{z^2}$ and $d_{xz}/d_{yz}$ with lobes pointing along the $z$-direction (i.e. the $a$-axis) are empty, which can mimic the $d^0$ effect for FE polarization along the $a$-axis. The orbital-projected electron density of state (PDOS) of VOF$_2$ monolayer is displayed in Fig. 2(a). Indeed, the topmost valence band near the Fermi energy level is from orbital $d_{xy}$, in consistent with the occupied charge density presented in Fig. 1(c). And VOF$_2$ monolayer is a semiconductor, with a band gap 0.6 eV in pure GGA-PBE calculation, which is usually underestimated.

Furthermore, the phonon spectrum of VOF$_2$ monolayer is calculated to verify the dynamic stability. As shown in Fig. 2(b), no obvious imaginary frequency mode is evidenced for the optimized FE state, implying that its FE monolayer is dynamically stable. The cleavage energy ($E_{cl}$) is also calculated.[40] By increasing the separation distance $d$ and keeping a 20 Å vacuum layer, the exfoliation process is simulated, as shown in Fig. 2(c). Comparing with the $E_{cl}$ of graphite (0.37 J/m$^2$), VOF$_2$ owns a smaller value (0.17 J/m$^2$) due to its weaker interlayer vdW interaction, close to the one of VOCl$_2$ monolayer (0.16 J/m$^2$).[28]

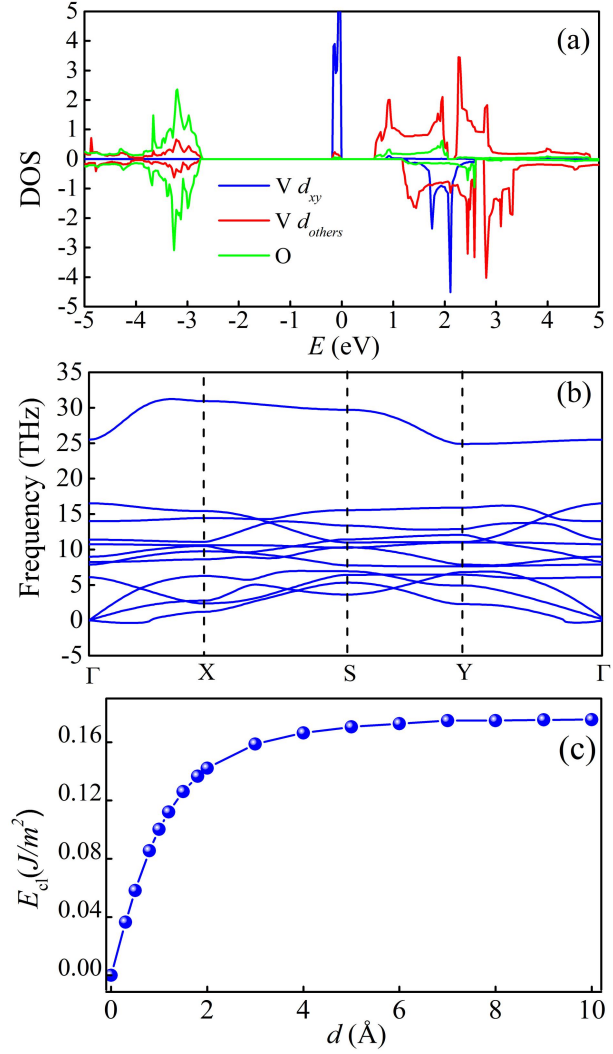

**Figure 2.** (a) The PDOS of VOF$_2$ monolayer. It is clear that the topmost valence band is from V's $d_{xy}$ orbital, while all other V's orbitals form the conducting bands. (b) The phonon spectra of VOF$_2$ monolayer. (c) The cleavage energy $E_{cl}$ as function of separation distance $d$ in the process of exfoliating VOF$_2$ monolayer from its bulk.

**Ferroelectricity & magnetism**

Subsequently, we will study the ferroelectricty and magnetism of VOF$_2$ monolayer. Obviously, the displacement of V ion from the center of octahedron, breaks the inversion symmetry and thus generates a spontaneous polarization. In our DFT calculation, the FE polarization of VOF$_2$ monolayer is estimated as 332 $p$C/m, larger than other VO$X_2$, as compared in Fig. 3(a). The polarization reversal process can also be simulated by choosing a appropriate switching path, i.e. by shifting the position of V, as illustrated in Fig. 3(b). If the lattice constants $a$ and $b$ of FE phase are fixed during this process, the energy barrier for FE switching is 0.33 eV. However, if the lattice constants $a$ and $b$ can adapt coordinately during the switching, the actual energy barrier for VOF$_2$ monolayer reduces to 0.15 eV, smaller than that of VOCl$_2$.[23] This barrier can be considered as an upper limit for real materials.

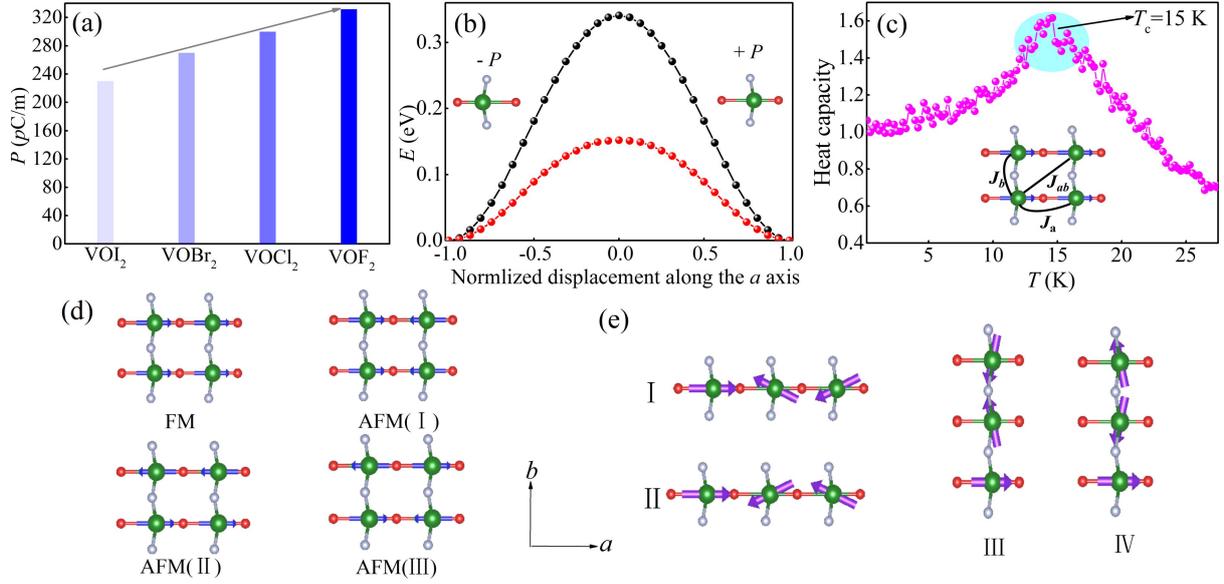

**Figure 3.** (a) Comparison of calculated polarization of VO$X_2$ ($X$=I, Br, Cl, and F). (b) Energy barrier of for FE switching, as a function of FE displacement of V ion. Black and red curves: the lattice constants ($a$ and $b$) are unchanged and changed during the switching, respectively. Inserts: +$P$ and -$P$ denote the polarization directions (c) The MC simulated heat capacity as a function of temperature ($T$) for VOF$_2$ monolayer. Insert: the exchange paths between V ions. (d-e) Schematic of magnetic orders considered in our DFT calculation. (d) Four types of collinear magnetic orders: FM, AFM(I), AFM(II), and AFM(III). (e) Four types of 120° non-collinear magnetic orders.

The magnetism comes from the spin of unpaired electron on the $d_{xy}$ orbital of V$^{4+}$ ion, and thus the local magnetic moment is about 1 μ$_B$/V. In order to study the magnetic ground state of VOF$_2$, a 2×2×1 supercell is adopted, in which four collinear magnetic orders can be defined: FM, AFM(I), AFM(II), and AFM(III), as shown in Fig. 3(d). The energy differences of these magnetic orders obtained in our DFT calculation are shown in Table II, indicating that the FM order is more stable than other AFM orders.

**Table II.** DFT energy difference of four types of magnetic orders and the energy of FM order (in unit of meV).

| $\Delta E_{FM}$ | $\Delta E_{AFM(I)}$ | $\Delta E_{AFM(II)}$ | $\Delta E_{AFM(III)}$ |
|---|---|---|---|
| 0 | 6 | 36 | 31 |

Then the magnetic exchanges, including the nearest-neighbor $J_a$ and $J_b$ along the $a$- and $b$-direction respectively and next-nearest-neighbor $J_{ab}$, are extracted using the following

equations:

$$E_{FM}=E_0+4J_a+4J_b+8J_{ab}, \quad E_{AFM(I)}=E_0-4J_a+4J_b-8J_{ab},$$

$$E_{AFM(II)}=E_0+4J_a-4J_b-8J_{ab}, \quad E_{AFM(III)}=E_0-4J_a-4J_b+8J_{ab}, \quad (1)$$

where $E_0$ denotes the energy of the nonmagnetic part and all energies are recalculated based on the optimized lattice with FM state. The calcualted values of exchanges are shown in Table III. All three exchanges tend to form FM coupling, and the exchange $J_b$ plays a leading role, which is reasonable considering the $d_{xy}$ orbital shape.

Besides the exchanges, the magnetic anisotropy is essential to stabilize a long range magnetic order in the 2D limit, because of the Mermin-Wagner theory.[41] By considering the SOC coupling, the magnetic anisotropy is obtained by calculating the energy of the FM orders with spin pointing along [100], [010], and [001] directions. The calculated magnetic anisotropy energy are 0.02 meV ($K_b$) and 0.05 meV ($K_c$), as shown in Table III, implying a magnetic easy axis along the $a$-axis.

**Table III.** Nearest-neighbor exchanges along the $a$- and $b$- directions ($J_a$ and $J_b$), next-nearest-neighbor exchange ($J_{ab}$), magnetic anisotropic coefficient, Dzyaloshinskii-Moriya vectors, extracted from DFT energies. The spin is normalized to unit one. Those components below 0.01 meV are set as zero for $D_a$ and $D_b$

| $J_a$ | $J_b$ | $J_{ab}$ | $K_b$ | $K_c$ | $D_a$ | $D_b$ |
|---|---|---|---|---|---|---|
| -0.06 | -3.81 | -0.34 | 0.02 | 0.05 | (0, 0, 0) | (0, 0, 0.09) |

Besides, the anti-symmetric Dzyaloshinskii-Moriya interaction may also affect the magnetic ground state.[42,43] For example, for VOI$_2$, although a FM ground state was predicted,[26] our following work suggests that the strong Dzyaloshinskii-Moriya interaction distorts the spin texture to a spiral one, instead of a collinear FM one. This Dzyaloshinskii-Moriya interaction directly associates with the FE distortion. To calculate the Dzyaloshinskii-Moriya coefficients, the 3×1×1 and 1×3×1 supercells are constructed and the non-collinear spin angels between neighboring V ions are set as 120°, as shown in Fig. 3(e). The 120° non-collinear configurations can converge well in our calculation, and the output magnetic textures are corrected as expected. Considering the crystal symmetry, only the $a$-component of $D_a$ and $c$-component of $D_b$ are allowed to be nonzero. Even though, the $a$-component of $D_a$ remains too small in DFT calculation, and the $c$-component of $D_b$ is estimate as 0.09 meV, as summarized in Table III. The weak magnetocrystalline anisotropy and Dzyaloshinskii-Moriya interaction are reasonably tiny, because here only light elements vanadium, oxygen, and fluorine present. Thus, it is safe to conclude that VOF$_2$'s ground state is FM.

In principle, the DFT calculation works at zero-temperature without thermal effect, while the MC simulation can deal with the temperature-dependent phase transition. Therefore, using above exchange interactions, magnetic anisotropic coefficient, and Dzyaloshinskii-Moriya interaction, a MC simulations on the Heisenberg spin model is performed to trace the temperature-dependence of magnetism. The spin model Hamiltonian is written as:

$$H = J_a \sum_{<i,j>_a} S_i \cdot S_j + J_b \sum_{<m,n>_b} S_m \cdot S_n + J_{ab} \sum_{\langle\langle q,w \rangle\rangle} S_q \cdot S_w + \sum_i [K_b(S_i^y)^2 + K_c(S_i^z)^2] \\ + D_a \cdot \sum_{<i,j>_a} S_i \times S_j + D_b \cdot \sum_{<m,n>_b} S_m \times S_n \quad , \quad (1)$$

where $S_i$ is the normalized vector of spin (i.e. $|S|=1$) at site $i$; $< >_{a/b}$ represents the nearest neighbor along the $a$-/$b$- direction; $\langle\langle \rangle\rangle$ represents the next nearest neighbor along the diagonal direction. $K_b$ and $K_c$ shows the magnetic anisotropy coefficient in/out of the plane. $D_a$ and $D_b$ represent antisymmetric Dzyaloshinskii-Moriya interaction along $a$-/b- direction. According to the specific heat curve shown in Fig. 3(c), a peak occurs around 15 K, suggesting a magnetic phase transition. The MC snapshot (not shown here) also confirms the FM texture. Last, we have to admit that the parameters used in MC simulation are based on the zero-temperature DFT calculation, which is qualitatively reasonable.

**The effect of additional $U$**

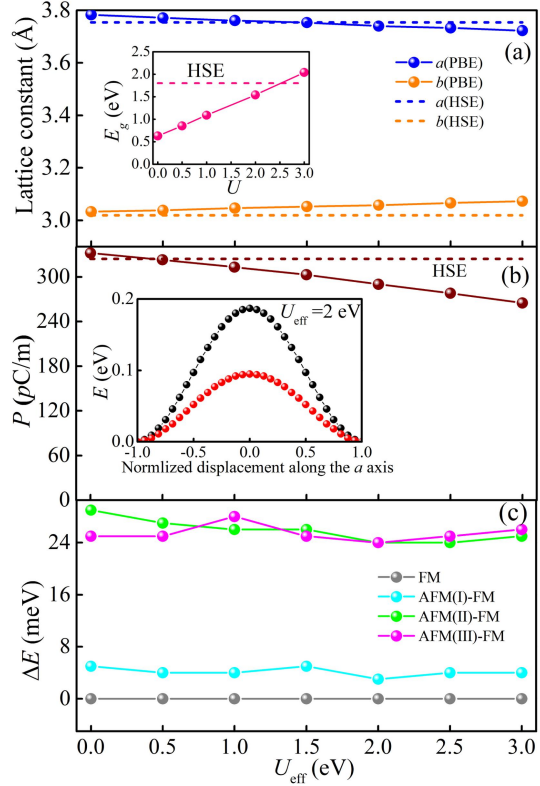

**Figure 4.** (a) Lattice constants $a$ and $b$, (b) FE polarization, and (c) energy differences of magnetic states as a function of of $U_{eff}$. The HSE06 results are also presented as broken lines for comparison. Insert of (a): Band gap. Insert of (b): Energy barrier for FE switching when $U_{eff}$ is 2 eV. Black and red curves correspond to the cases of lattice constants unchanged and changed during the process.

Based on pure spin-polarized GGA calculation, above results have revealed that $VOF_2$ monolayer is a rare intrinsic 2D FE and FM material. Even through, it is essential to double check the influence of Hubbard $U$ correction, considering the valence electron is from a $3d$ orbital. Then, the GGA+$U$ method is adopted to re-check the structure, ferroelectric and magnetic properties of $VOF_2$ monolayer.

As shown in Fig. 4(a), the lattice constants $a$ and $b$ do change obviously with increasing $U_{eff}$ value, e.g. within $\pm 2.5\%$. And the band gap increases from original 0.6 eV to 2.1 eV (when $U_{eff}$=3 eV). Meanwhile, the FE polarization reduces from 332 $pC/m$ to 264 $pC/m$, as shown in Fig. 4(b). The physical reason is that the increased $U_{eff}$ suppresses the $d$-$p$ orbital hybridization between V and O, i.e. the driving force of its proper ferroelectricity. The energy barrier for FE switching also decreases with increasing $U_{eff}$, e.g. 0.095 eV when $U_{eff}$=2 eV. We also re-calculate the energies of four magnetic orders. Despite the value of $U_{eff}$, the energies of AFM configurations are always higher than the FM one, as shown the Fig. 4(c).

Thus it is safe to conclude that magnetic ground state of $VOF_2$ monolayer is FM and FE, which will not be altered by the choice of Hubbard $U$ in DFT calculation.

**Strain effect**

Comparing with 3D crystals, one advantage of 2D materials is their structural flexibility. The magnetic and electronic structures may be tunable by stress.[44] To investigate this effect, we apply the unaxial compressive/tensile strain from -5% to +5% along the $b$-axis, then the lattice constant along the $a$-axis and ions' positions are optimized accordingly. As shown in Fig. 5(a), the lattice constant $a$ changes almost linearly as a function of strain, in opposite to the tendency of $b$-axis but with a relative smaller magnitude, i.e. from +1.3% to -0.8%. In other words, the stiffness along the $b$-axis is softer and that along the $a$-axis is harder.

Figure 5(b) depicts the strain dependent FE polarization, which decreases with the decreasing lattice constant $a$, as expected. Or in other words, the compressive strain along the $b$-axis can enhance the FE polarization.

Most interestingly, there is competition between the FM order and AFM(III) order. The AFM(III) state owns a lower energy than the FM one when compressive strain goes beyond

-1%, implying a magnetic phase transition for VOF$_2$ monolayer.

In short, the compressive strain along the *b*-axis (or tensile strain along the *a*-axis) is an effective method to tune the multiferroicity of VOF$_2$ monolayer: not only to enhance its FE polarization but also to change the magnetic ground state.

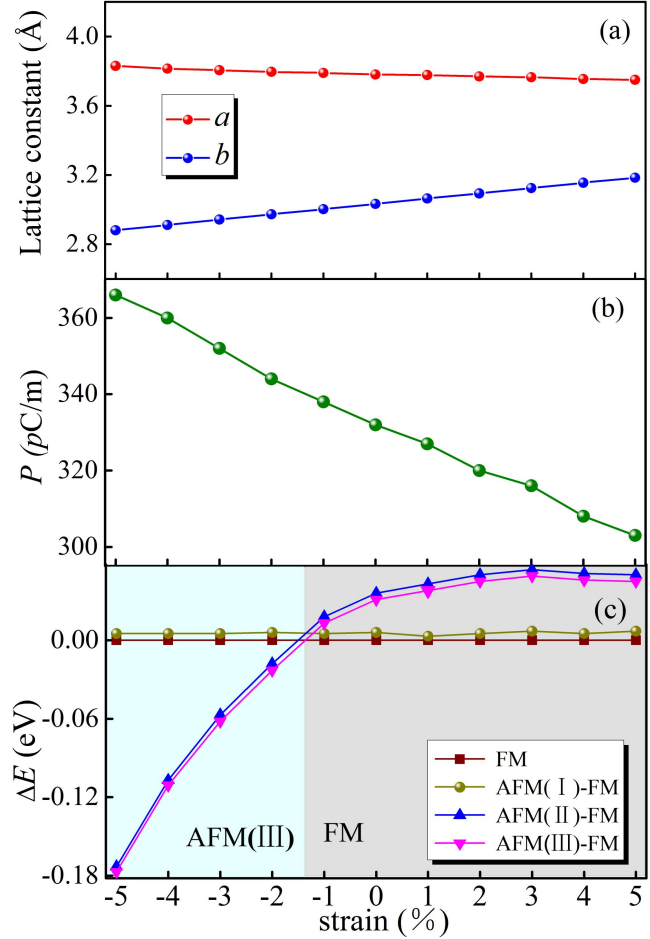

**Figure 5**. (a) Lattice constants *a* and *b*, (b) ferroelecric polarization, and (c) energy differences of magnetic states as a function of of unaxial strain applied along the *b*-direction. The magnetic ground state changes from the FM one in the unstrained condition to the AFM(III) one under a moderate compressive strain.

## Conclusion

In summary, we have performed a systematic DFT calculation on a 2D oxyhalide VOF$_2$ monolayer, covering its crystal structure, electronic properties, structural stability, FE polarization, as well as its magnetism. Our results suggest that VOF$_2$ is an intrinsic 2D multiferroic material with rare coexistence of ferromagnetism and ferroelectricity. Furthermore, the unaxial strain can effectively tune its FE polarization and magnetic ground state. Our theoretical work predicts a potential 2D multiferroic material with highly desired properties, which needs further experiment verification.

■ ACKNOWLEDGMENTS

This work was supported by the National Natural Science Foundation of China (Grants Nos. 11834002 and 11674055). Computing resources used in this work were mainly provided by the Big Data Center of Southeast University.